\documentclass[aps,twocolumn,pra,superscriptaddress,showpacs,tightenlines,nofootinbib]{revtex4}

\usepackage{graphicx,times}
\usepackage{amstext}
\usepackage{amsmath}            
\usepackage{amssymb}            
\usepackage{graphicx}             
\usepackage{latexsym}
\usepackage{bm}

\def\extra#1{{``#1''}}

\newcommand{\ER}[1]{{E_{R}^{\rm (#1)}}}
\newcommand{\RHO}[1]{{\rho^{\rm (#1)}}}
\newcommand{\SIGMA}[1]{{\sigma^{(\rm #1)}}}

\def\<{\langle}
\def\>{\rangle}

\newcommand{\ket}[1]{|#1\rangle}

\newcommand{\jpa}{J. Phys. A~}

\newcommand{\pla}{Phys. Lett. A~}

\begin{document}

\title{Two-qubit mixed states more entangled than pure states:\\
Comparison of the relative entropy of entanglement for a given
nonlocality}

\author{Bohdan Horst}

\affiliation{Faculty of Physics, Adam
 Mickiewicz University, 61-614 Pozna\'n, Poland}

\author{Karol Bartkiewicz}
\email{bartkiewicz@jointlab.upol.cz} \affiliation{RCPTM, Joint
Laboratory of Optics of Palack\'y University and Institute of
Physics of Academy of Sciences of the Czech Republic, 17.
listopadu 12, 772 07 Olomouc, Czech Republic }

\author{Adam Miranowicz}

\affiliation{Faculty of Physics, Adam
 Mickiewicz University, 61-614 Pozna\'n, Poland}

\date{\today}

\begin{abstract}
Amplitude damping  changes entangled pure states into usually
less-entangled mixed states. We show, however, that even
local amplitude damping of one or two qubits can result in
mixed states more entangled than pure states if one compares
the relative entropy of entanglement (REE) for a given degree
of the Bell-Clauser-Horne-Shimony-Holt inequality violation
(referred to as nonlocality). By applying Monte-Carlo
simulations, we find the maximally entangled mixed states and
show that they are likely to be optimal by checking the
Karush-Kuhn-Tucker conditions, which generalize the method of
Lagrange multipliers for this nonlinear optimization problem.
We show that the REE for mixed states can exceed that of pure
states if the nonlocality is in the range (0,0.82) and the
maximal difference between these REEs is 0.4. A former
comparison [Phys. Rev. A {\bf 78}, 052308 (2008)] of the REE
for a given negativity showed analogous property but the
corresponding maximal difference in the REEs is one-order
smaller (i.e., 0.039) and the negativity range is (0,0.53)
only. For appropriate comparison, we normalized the
nonlocality measure to be equal to the standard entanglement
measures, including the negativity, for arbitrary two-qubit
pure states. We also analyze the influence of the
phase-damping channel on the entanglement of the initially
pure states. We show that the minimum of the REE for a given
nonlocality can be achieved by this channel, contrary to the
amplitude damping channel.
\end{abstract}



\maketitle \pagenumbering{arabic}

\section{Introduction}

Quantum entanglement and nonlocality~\cite{Bell,Clauser}, which
are two special aspects of quantum correlations, are among the
central concepts in quantum theory providing powerful resources
for modern quantum-information processing~\cite{Nielsen}.
Nevertheless, despite much progress (as reviewed in, e.g.,
Ref.~\cite{Horodecki09}), our understanding of the relation
between entanglement and nonlocality is very incomplete.

Quantum nonlocality is often discussed in the context of Bell's
theorem, which can arguably be considered ,,the most profound
discovery of science''~\cite{Stapp75}. The Bell inequality
violation (BIV) for a given state implies that there is no
physical theory of local-hidden variables which can reproduce the
predictions for the state~\cite{Bell}. Although the BIV does not
necessarily imply nonlocality and vice versa (see, e.g.,
Ref.~\cite{Popescu94}), for convenience, we use the terms BIV and
nonlocality interchangeably.

Because of the prominent role of quantum entanglement, much
effort has been devoted to investigating measures of
entanglement that can be used for its quantification.
Currently, a variety of formal and operational entanglement
measures can be applied~\cite{Horodecki09} depending on the
physical and information contexts. For example, if one is
interested in distinguishing entangled from separable states,
a convenient choice is the relative entropy of entanglement
(REE)~\cite{Vedral97a}. However, if one considers the cost of
entanglement under operations preserving the positivity of
partial transpose (PPT)~\cite{Audenaert}, the appropriate
choice of entanglement measure is the (logarithmic)
negativity~\cite{Peres,Horodecki}. The
concurrence~\cite{Wootters98} is yet another popular measure
of entanglement, which quantifies the entanglement of
formation~\cite{Bennett1}.

For two-qubit pure states, entanglement and nonlocality are
equivalent resources as shown by Gisin~\cite{Gisin91}. The
interest in comparisons of BIV and entanglement was arguably
triggered by the discovery of two-qubit entangled mixed
states (now called the Werner states) admitting a
hidden-variable model~\cite{Werner89}. For mixed states, not
only the entanglement and nonlocality become inequivalent
phenomena, but even the measures of entanglement become, in
general, inequivalent. For example, two-qubit mixed states of
the same value of a given entanglement measure can correspond
to different values of entanglement quantified by other
measures. Thus, it can happen that entangled states ordered
according to one entanglement measure can be ordered
differently by another
measure~\cite{Eisert99,Miran04c,Miran04a,Miran04b}.

The most interesting entangled states are, arguably, the boundary
states, which are extremal in one measure for a given value of
another measure. They can serve as ``waymarks'' for entangled
states providing means for their systematic classification.
Moreover, it turns out that some of the most entangled states
according to the REE for a given value of another entanglement
measure as, e.g., the negativity~\cite{Miran08b}, are mixed
states.

In this paper, we find maximally entangled mixed states
(MEMS) corresponding to the largest REE for a given degree of
nonlocality using the Horodecki
measure~\cite{Horodecki95,Horodecki96}. We also show that
these MEMS are more entangled than pure states for a wide
range of this nonlocality degree. As an indicator of
extremality of the found MEMS we use the Karush-Kuhn-Tucker
(KKT) conditions in a refined method of Lagrange multipliers.

The Horodecki measure $B(\rho)$ of BIV~\cite{Horodecki95} for
two-qubit states $\rho$ enabled detailed quantitative
comparisons with other measures of quantumness including
entanglement. The work of Verstraete and
Wolf~\cite{Verstraete02}, where the upper and lower bounds of
$B(\rho)$ for a given value of the concurrence $C(\rho)$ were
found, is especially relevant to the present paper, where we
find the bounds of $B(\rho)$ for a given value of the REE,
$E_R(\rho)$, for \emph{arbitrary} two-qubit states. Other
studies were usually limited to \emph{specific} classes of
two-qubit states generated in some physical processes
(including decoherence). For example, the nonlocality
measures were such compared with: the concurrence $C(\rho)$
(or, equivalently, the entropy of
formation)~\cite{Jakobczyk03,Miran04c,Derkacz04,
Liao07,Kofman08,Deng09, Mazzola10,Berrada11,Hu13}, negativity
$N(\rho)$~\cite{Miran04c}, fidelity
$F(\rho)$~\cite{Kowalewska06,Hu13}, as well as the purity
$P(\rho) ={\rm \mathrm{Tr}} (\rho^2)$~\cite{Deng09,
Mazzola10} and the closely related degrees of mixedness as
measured by either their linear entropy $S_L(\rho)=
1-P(\rho)$~\cite{Derkacz04,Deng09} or participation ratio
$R(\rho) = 1/P(\rho)$~\cite{Batle11}.

The paper is organized as follows. In
Sec.~\ref{sec:definitons}, we introduce some basic
definitions of the nonlocality and entanglement measures used
in our paper. In Sec.~\ref{sec:special_states}, we
demonstrate analytically that there are mixed states more
entangled than pure states when analyzing the REE for a fixed
nonlocality (and negativity). In Sec.~\ref{sec:damping}, we
describe how to physically obtain an important class of
entangled states, which, as we show in Sec.~\ref{sec:KKT},
are the boundary states if quantified by the REE for given
values of the nonlocality, negativity, and concurrence. The
most important result in this paper is finding the states
exhibiting the highest and lowest REEs for a given value of
the nonlocality and demonstrating that mixed states can be
more entangled than pure states if the fixed nonlocality is
less then $0.8169$.

\section{Definitions}
\label{sec:definitons}

Hereafter, we study general two-qubit density matrices $\rho$,
which can be expressed in the standard Bloch representation as
\begin{equation}
\rho  =  \frac{1}{4}(I\otimes I+\vec{x}\cdot\vec{\sigma}\otimes
I+I\otimes\vec{y}\cdot\vec{\sigma}+\!\!\!\sum
\limits_{n,m=1}^{3}T_{nm}\,\sigma _{n}\otimes \sigma _{m}),
\label{rho}
\end{equation}
where $\vec{\sigma}=[\sigma_{1},\sigma_{2},\sigma_{3}]$ and the
correlation matrix
$T_{ij}=\mathrm{Tr}[\rho(\sigma_{i}\otimes\sigma_{j})]$ are given
in terms of the three Pauli matrices, and
$x_{i}=\mathrm{Tr[}\rho(\sigma_{i}\otimes I)]$
($y_{i}=\mathrm{Tr[}\rho(I\otimes\sigma_{i})]$) are the elements
of the Bloch vector $\vec{x}$ ($\vec{y}$) of the first (second)
subsystem. Expressing the two-qubit density matrix by
Eq.~(\ref{rho}) is very convenient since it allows a direct
application of an effective criterion for the nonlocality.

\subsection{Nonlocality measure}

The two-qubit Bell inequality in the form derived by Clauser, Horne,
Shimony and Holt (CHSH)~\cite{Clauser} can be formulated as
\begin{equation}
|\mathrm{Tr}\,(\rho \,{\mathcal B}_{\mathrm{CHSH}})|\leq 2,
\label{CHSH}
\end{equation}
where the Bell-CHSH operator is ${\mathcal B}_{\mathrm{CHSH}}$ is
given by
\begin{equation}
{\mathcal B}_{\mathrm{CHSH}}=\vec{a}\cdot \vec{\sigma }\otimes
(\vec{ b}+\vec{b}^{\prime })\cdot \vec{\sigma }+\vec{a}^{\prime
}\cdot \vec{\sigma }\otimes (\vec{b}-\vec{b}^{\prime })\cdot
\vec{\sigma }, \label{BellOp}
\end{equation}
and its expected value is maximized over real-valued
three-dimensional unit vectors $\vec{a},\,\vec{a}^{\prime
},\,\vec{b},$ and $\vec{b}^{\prime }$. According to the
Horodecki theorem, the maximum expected value of the
Bell-CHSH operator for a given state $\rho$ reads
as~\cite{Horodecki95,Horodecki96}:
\begin{equation}
\max_{{\mathcal B}_{\mathrm{CHSH}}}\,\mathrm{Tr}\,(\rho
\,{\mathcal B}_{ \mathrm{CHSH}})=2\,\sqrt{M(\rho )}
\label{MaxBellOp}
\end{equation}
given in terms of the parameter
\begin{equation}
M(\rho )=\max_{j<k}\;\{h_{j}+h_{k}\}\leq 2, \label{M}
\end{equation}
where $h_{j}$ $(\,j=1,2,3)$ are the eigenvalues of the real
symmetric matrix $U=T^{T}\,T$ constructed from the
correlation matrix $T$ and its transpose $T^{T}$. Hence, the
condition for violating the Bell-CHSH inequality is $M(\rho
)>1$~\cite{Horodecki95,Horodecki96}. For convenience, we
refer to \emph{nonlocality} as the violation of the Bell-CHSH
inequality.

To quantify a degree of the nonlocality, one can directly use
$M(\rho)$ or $2\,\sqrt{M(\rho)}$ (see, e.g.,
Refs.~\cite{Ghosh,Jakob}), or more naturally $\max
\,[0,\,M(\rho )-1\,]$ (see, e.g., Ref.~\cite{Jakobczyk03}).
However, we decided to use another function of $M(\rho)$,
denoted by $B(\rho)$, which for two-qubit pure states is
equal to the concurrence and negativity. This measure can be
given as~\cite{Miran04b}:
\begin{equation}
B(\rho )\equiv \sqrt{ \max \,[0,\,M(\rho )-1]}. \label{BIV}
\end{equation}
We see that $B(\rho)=0$ if a given state $\rho$ satisfies the
Bell-CHSH inequality, given by Eq.~(\ref{CHSH}), and $B(\rho)=1$
if the inequality is maximally violated. The value of $B>0$
increases with $M$, thus it can be used to quantify the BIV. We
refer to this BIV degree as the nonlocality measure.

\subsection{Entanglement measures}

Now, we recall some definitions of a few selected measures of
entanglement applicable for two-qubit entangled states.

In our considerations, the most important entanglement
measure is the REE defined as $E_R(\rho)={\rm min}_{\sigma\in
{\cal D}} S(\rho ||\sigma)$, which is the relative entropy
$S(\rho ||\sigma )={\rm Tr}\,( \rho \log_2 \rho -\rho\log_2
\sigma)$ minimized over the set ${\cal D}$ of separable
states $\sigma$~\cite{Vedral97a,Vedral98}. This measure is a
quantum counterpart of the Kullback-Leibler divergence
quantifying the difference between two classical probability
distributions. Evidently, the REE is limited, by definition,
to distinguishing a density matrix $\rho$ from the closest
separable state (CSS) $\sigma$ only. Note that the REE is not
a true metric, since it is not symmetric and does not fulfill
the triangle inequality. However, the REE has a desirable
property of a good entanglement measure that, for pure
states, it reduces to the von Neumann entropy of one of the
subsystems.

Unfortunately, as discussed in
Refs.~\cite{Eisert05,Ishizaka03,Miran08a,Kim10}, it is very
unlikely to find an analytical compact formula for the REE of a
general two-qubit mixed state, which would correspond to finding
its CSS. Numerical procedures for calculating the two-qubit REE
correspond usually to an optimization problem over 79 or more real
parameters~\cite{Vedral98,Miran08b,Zinchenko10}. On the other
hand, there is a compact-form solution of the inverse problem: If
a CSS is known then all the entangled states (having the same CSS)
can be given analytically not only for two
qubits~\cite{Ishizaka03,Miran08a} but even for arbitrary
multipartite states of any dimensions~\cite{Friedland11}.

The second measure studied here is the negativity
\cite{Zyczkowski98,Eisert99,Vidal}, defined as ${N}({\rho})=\max
\{0,-2\mu _{\min}\}$, where $\mu_{\min}=\min{\rm
eig}(\rho^{\Gamma})$ and $\Gamma$ denotes partial transpose. The
negativity is related to the logarithmic negativity, $\log_2[{N}
({\rho})+1]$, which is a measure of the entanglement cost under
the PPT operations~\cite{Audenaert,Ishizaka04}. These two related
measures reach unity for the Bell states and vanish for separable
states, however, for clarity of our further presentation, we will
use only the negativity.

The last entanglement measure applied in this paper is the
concurrence introduced by Wootters~\cite{Wootters98} as
$C({\rho})=\max \{0,2\lambda_{\max}-\sum_j\lambda_j\}$, where
$\lambda^2 _{j} = \mathrm{eig}[{\rho }({\sigma }_{2}\otimes
{\sigma }_{2}){\rho}^{\ast }({ \rho }_{2}\otimes {\sigma
}_{2})]_j$ and $\lambda_{\max}=\max_j\lambda_j$. This measure is
directly related to the entanglement of formation,
$E_{F}({\rho})$~\cite{Bennett1}. However, for the same reason as
in the case of the negativity we use the concurrence instead of
$E_{F}({\rho})$.

\section{Analytical comparison of entanglement and nonlocality for SPECIAL states}
\label{sec:special_states}

In Fig.~\ref{fig:1}, we presented several curves corresponding, in
particular, to the Horodecki states, Bell-diagonal states, and
pure states. The REE for these states can be calculated
analytically so let us first discuss and compare them to
demonstrate the main point of our paper.

\subsection{Pure states}

We can simply relate all the above-mentioned entanglement and
nonlocality measures in a special case of an arbitrary two-qubit
pure states
$|\psi\rangle=a|00\rangle+b|01\rangle+c|10\rangle+d|11\rangle$
(with the normalization condition $|a|^2+|b|^2+|c|^2+|d|^2=1$) as
\begin{eqnarray}
B(|\psi\rangle)=C(|\psi\rangle)=N(|\psi\rangle)=2|ad-bc|.
\label{B_pure_state}
\end{eqnarray}
Moreover, in this special case $B$, $N$ and $C$ are simply related
to the REE and von Neumann's entropy $S$ as\begin{eqnarray} {\cal
W}(B) &=& E_{R}(|\psi\rangle)=S(\rho_i), \label{N35b}
\end{eqnarray}
where ${\cal W}(B) \equiv h\left(\frac{1}{2}
[1+\sqrt{1-B^2}]\right)$ is the Wootters
function~\cite{Wootters98} given in terms of the binary
entropy $h(x)=-x\log _{2}x-(1-x)\log _{2}(1-x)$, and
$\rho_i=\mathrm{Tr}_{3-i}\rho$ is the reduced density matrix
of the $i$th qubit ($i=1,2$).

\subsection{Horodecki states}

Let us also analyze a mixture of a Bell state, say
$|\psi^{+}\rangle=(|01\rangle+|10\rangle)/\sqrt{2}$, and vacuum
state, i.e.,
\begin{equation}
\RHO{H}(p)=p|\psi^{+}\rangle\langle\psi^{+}|
+(1-p)|00\rangle\langle00|,\label{rho_H}
\end{equation}
which is  referred to as the Horodecki state. By applying
Eq.~(\ref{M}), we find that
$M(\RHO{H})=p^{2}+\max[p^{2},(1-2p)^{2}]$. Thus, the nonlocality
$B$ is given by
\begin{equation}
B(\RHO{H})=\sqrt{\max(0,2p^{2}-1)}. \label{rhoH_B}
\end{equation}
On the other hand, the entanglement measures are the
following: the concurrence is $C(\RHO{H})=p$, the negativity is
$N(\RHO{H})  =  \sqrt{(1-p)^{2}+p^{2}}-(1-p),$ and the REE reads
as
\begin{eqnarray}
E_{R}(\RHO{H})  =  2h(1-p/2)-h(p)-p\hspace{1.7cm}\nonumber \\
  =  (p-2)\log_{2}(1-p/2)+(1-p)\log_{2}(1-p).\label{REE_H}
\end{eqnarray}
It is seen that the Horodecki state is entangled for $0<p\le1$,
while it is violating the Bell-CHSH inequality only for
$1/\sqrt{2}<p\le1$. By comparing the REEs for a given nonlocality
for the Horodecki and pure states we find that
\begin{eqnarray}
\ER{H}(B)>\ER{P}(B) & \quad & {\rm for}\;0<B<B_{6},\label{N14a}\\
\ER{H}(B)<\ER{P}(B) & \quad & {\rm for}\; B_{6}<B<1,\nonumber
\end{eqnarray}
where $B_{6}\equiv B(\rho_{6})=0.5856$ and
$\ER{H}(B_{6})=\ER{P}(B_{6})=0.4520$ [see
Table~\ref{tab:points} and Fig.~\ref{fig:3}(c)]. This means
that mixed states can be more entangled than pure states at
least for $B<B(\rho_{6})$. Figure~\ref{fig:1}(c) shows that
this property holds up to $B<B(\rho_{5})=0.8169$ but for
mixed states different from the Horodecki states, which is
demonstrated in Sec.~\ref{sec:KKT}. Here, $\ER{H}$ as a
function of $B$ is explicitly given by Eq.~(\ref{REE_H}) for
$p=\sqrt{(1+B^2)/2} \ge1/\sqrt{2}$.

Analogous comparison of the REEs for a given negativity $N$ for
the Horodecki and pure states [see Fig.~\ref{fig:1}(b)] shows
that~\cite{Miran08b}:
\begin{eqnarray}
\ER{H}(N)>\ER{P}(N) & \quad & {\rm for}\;0<N<N_1,\label{N15a}\\
\ER{H}(N)<\ER{P}(N) & \quad & {\rm for}\; N_1<N<1,\nonumber
\end{eqnarray}
where $N_1\equiv N(\rho_{1})=0.3770$ and
$\ER{H}(N_1)=\ER{P}(N_1)=0.2279$ [see Table~I and
Fig.~\ref{fig:3}(b)]). Note that $\ER{H}$ as a function of
$N$ can be given explicitly by Eq.~(\ref{REE_H}) for
$p=\sqrt{2N(1+N)}-N$.

It is convenient to introduce a parameter $\Delta E_R(X)$, which
is the maximal difference in the REE between a given mixed state
$\rho$ and some pure state $\RHO{P}=|\psi\rangle\langle\psi|$
having the same value of either the nonlocality ($X=B$) or
negativity $(X=N)$, i.e.,
\begin{eqnarray}
\Delta
E_R(X')&=&\max_{X}\big\{E_{R}[\rho(X)]-E_{R}[\RHO{P}(X)]\big\}
\nonumber\\
&=&E_{R}[\rho(X')]-E_{R}[\RHO{P}(X')], \label{Delta}
\end{eqnarray}
where $X'$ is the optimal value of $X$. For the Horodecki state,
we observe that this maximal difference for the nonlocality is
equal to $\Delta\ER{H}(B')=0.2949$, which occurs for $B'=0$, while
for the negativity is only $\Delta \ER{H}(N')=0.0391$, which is
for $N'=0.1540$.

We also note that the Horodecki states are the lower bound of
the REE vs concurrence as shown in Fig.
1(a)~\cite{Verstraete02}.

\subsection{Bell-diagonal states}

Finally, let us also analyze the Bell-diagonal states (labeled by
D), which are defined by
\begin{eqnarray}
\RHO{D}=\sum_{i=1}^{4}\lambda_{i}|\beta_{i}\rangle\langle\beta_{i}|,\label{N12}
\end{eqnarray}
where $|\beta_{i}\rangle$ are the Bell states and
$0<\lambda_{j}<1$ such that $\sum_{j}\lambda_{j}=1$ with the
largest eigenvalue $\max_{j}\lambda_{j}\equiv(1+N)/2\ge1/2$. The
nonlocality of the Bell-diagonal state $\RHO{D}$ is given
by~\cite{Miran04b}:
\begin{equation}
B(\RHO{D})\!=\!\!\sqrt{\max\{0,2\max_{(i,j,k)}[(\lambda_{i}-\lambda_{j})^{2}+(\lambda_{k}-\lambda_{4})^{2}]-1\}},
\end{equation}
where subscripts $(i,j,k)$ correspond to the cyclic permutations
of $(1,2,3)$. By contrast, the negativity and concurrence are the
same and simply given by  $N(\RHO{D})=C(\RHO{D})=N.$ The REE
versus the negativity (and, thus, also the concurrence) reads as
\begin{equation}
E(\RHO{D})=1-h\Big(\frac{1+N}{2}\Big)\label{REE_N}
\end{equation}
as given in terms of the binary entropy $h$. If
$\max_{j}\lambda_{j}\le1/2$ then the state is separable
$E(\RHO{N})=0$.

It is evident that the nonlocality of the Bell-diagonal states
depends on all probabilities $\lambda_{i}$, while the entanglement
measures depend solely on the largest value
$\max_{i}\lambda_{i}>1/2$. Nevertheless, in some special cases of
these states, the entanglement and nonlocality measures can be
equal. For example, when only two probabilities $\lambda_{i}$
corresponding to  $|\psi^{\pm}\rangle$ are nonzero, the
Bell-diagonal state $\RHO{D}$ reduces to
\begin{equation}
\RHO{D2}=p|\psi^{+}\>\<\psi^{+}|+(1-p)|\psi^{-}\>\<\psi^{-}|,
\label{rho_dc2}
\end{equation}
which is studied in a physical context in Sec.~\ref{sec:damping}.
The nonlocality for this rank-2 state is simply given by
\begin{equation}
B(\RHO{D2})=|2p-1|\label{B_D2}
\end{equation}
implying that
\begin{equation}
E(\RHO{D2})=1-h\Big(\frac{1+B}{2}\Big),\label{REE_B}
\end{equation}
which corresponds to Eq.~(\ref{REE_N}) .

It is seen in Figs.~\ref{fig:1}(b) and~\ref{fig:1}(c) (see also
Sec.~\ref{sec:KKT} for a partial proof) that the lower bounds of
the REE vs negativity and the REE vs nonlocality correspond to the
Bell-diagonal state $\RHO{D}$, which satisfy the extremal KKT
conditions as we show in Sec.~\ref{sec:KKT}. Their physical
context is discussed below.

\section{Manipulating entanglement and nonlocality via damping channels}
\label{sec:damping}

\begin{figure}
\includegraphics[width=6.5cm]{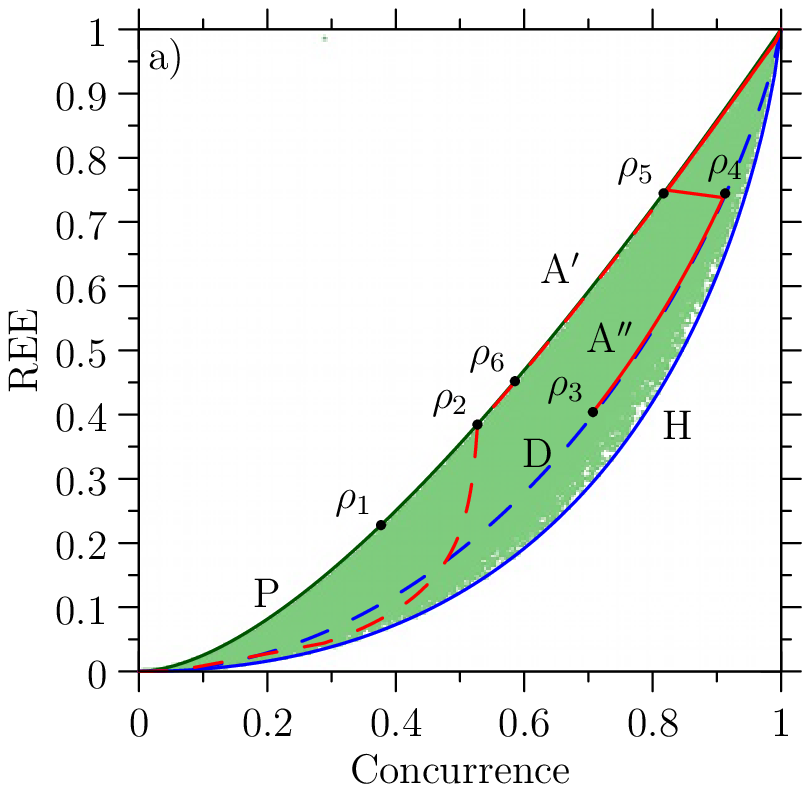}
\includegraphics[width=6.5cm]{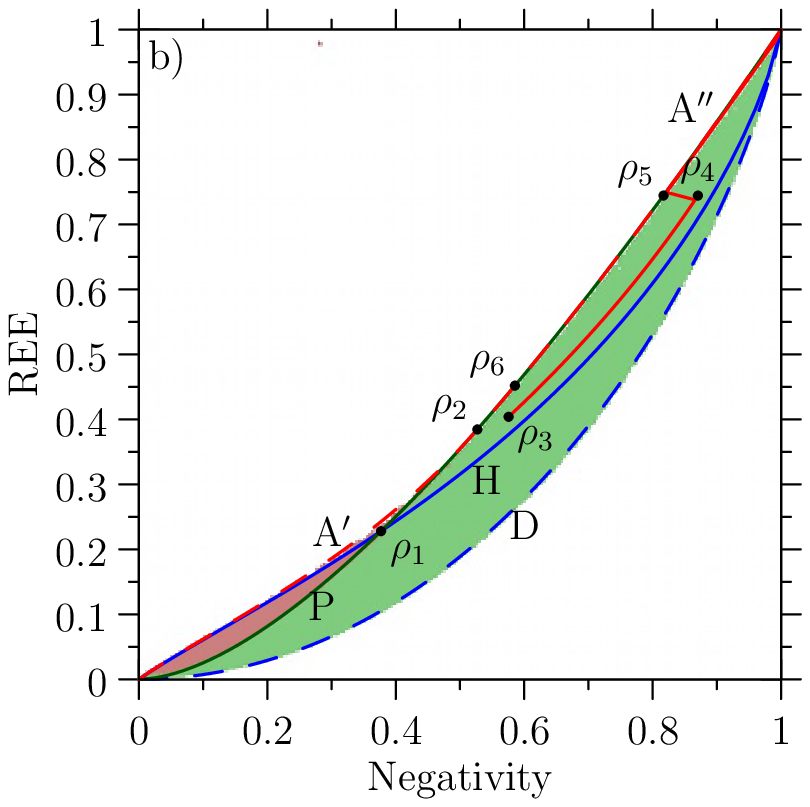}
\includegraphics[width=6.5cm]{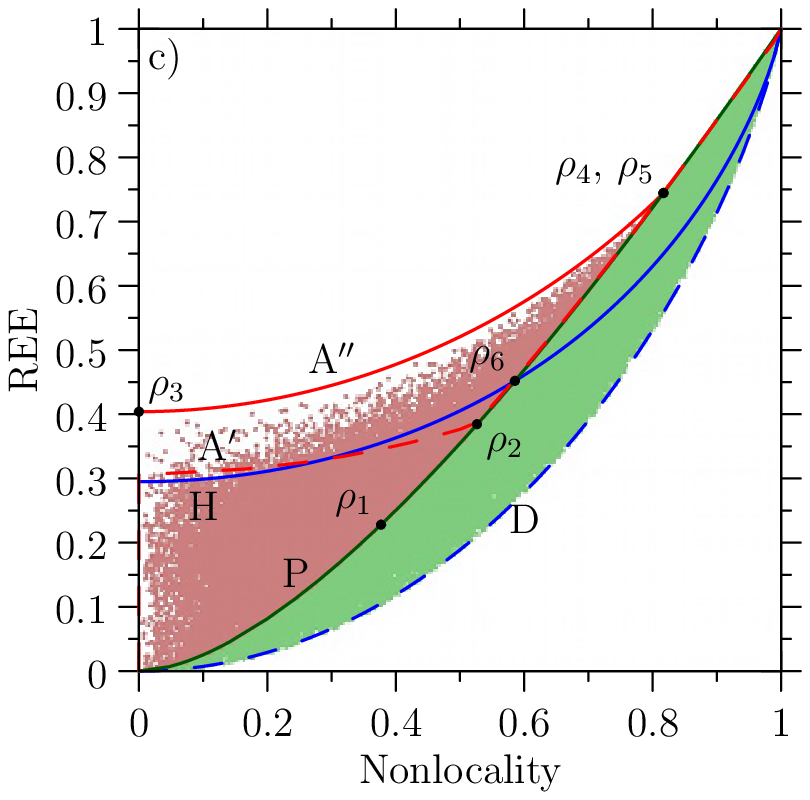}
\caption{\label{fig:1} (Color online)  Monte Carlo simulations of
about $10^6$ two-qubit states $\rho$ for the relative entropy of
entanglement, $E_R(\rho)$, versus: (a) concurrence $C(\rho)$, (b)
negativity $N(\rho)$, and (c) nonlocality $B(\rho)$ corresponding
to the Bell-CHSH inequality violation. Red regions correspond to
mixed states more entangled than pure states in terms of the
$E_R(\rho)$. Key: P, pure states, D, the Bell-diagonal states
($\alpha = 0.5$), A' (A''), the MEMS, which are the
amplitude-damped states optimized for the case b (c), H, the
Horodecki states, and the coordinates of points $\rho_n$ (where
$n=1,...,6$) are given in Table~\ref{tab:points}.}
\end{figure}
\begin{table}
\caption{\label{tab:points} Nonlocality and entanglement measures
of the amplitude-damped states $\rho_n=\RHO{A}(\alpha,p)$ (for
$n=1,...,6$), given by Eq.~(\ref{rho_adc}), corresponding to the
characteristic points marked in Fig.~\ref{fig:1}.}
\begin{ruledtabular}
\begin{tabular}{ccccccc}
State  & $\alpha$ & $p$ & $C$  & $N$ &  $E_R$ & $B$   \\
\hline
$\rho_1$& 0.0369& 1.0000  &  0.3770 &  0.3770 & 0.2279 &  0.3770  \\
$\rho_2$& 0.0751 & 1.0000  & 0.5271 & 0.5271 & 0.3847 & 0.5271  \\
$\rho_3$& 0.2198 & 0.8536  & 0.7070 & 0.5756 & 0.4039 & 0.0000  \\
$\rho_4$& 0.3510 & 0.9565  & 0.9130 & 0.8706 & 0.7445 & 0.8169  \\
$\rho_5$& 0.2116 & 1.0000  & 0.8169 & 0.8169 & 0.7445 & 0.8169  \\
$\rho_6$& 0.0947 & 1.0000  & 0.5856 & 0.5856 & 0.4520 & 0.5856  \\
\end{tabular}
\end{ruledtabular}
\end{table}
\begin{figure}
\includegraphics[width=6.5cm]{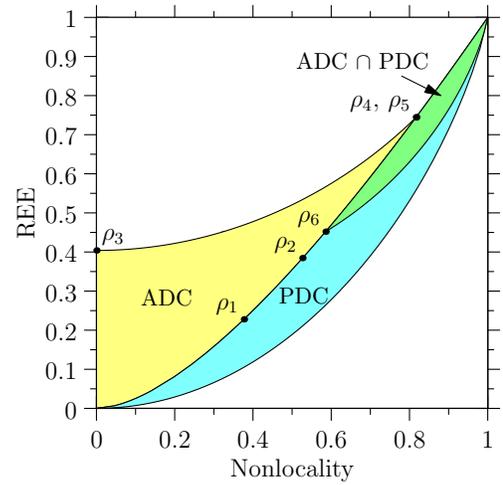}
\caption{\label{fig:2} (Color online) Ranges of the relative
entropy of entanglement for a given nonlocality for the two-qubit
mixed states generated from pure states by  the amplitude
damping channel, given by Eq.~(\ref{rho_adc}) (yellow and green
areas) and the phase damping channel, given by Eq.~(\ref{rho_pdc})
(blue and green areas). }
\end{figure}
\begin{figure}
\includegraphics[width=6.5cm]{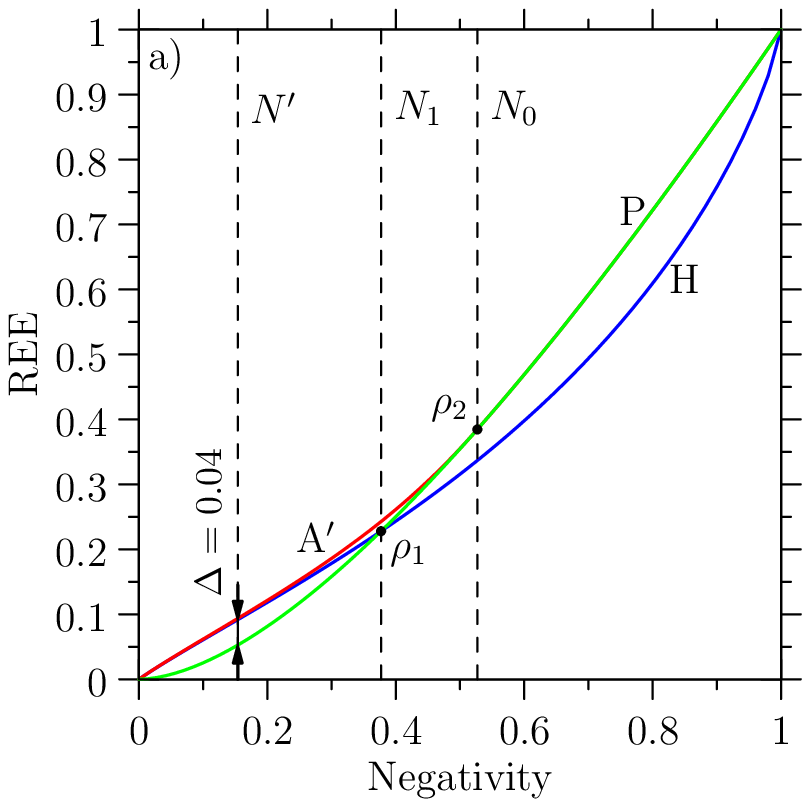}
\includegraphics[width=6.5cm]{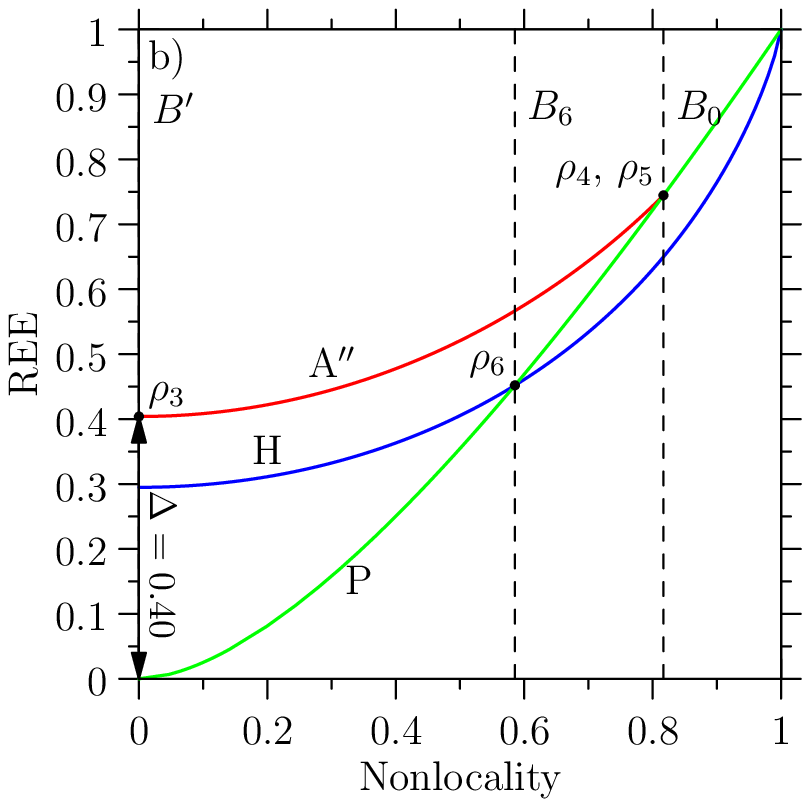}
\caption{\label{fig:3} (Color online)  Ranges of the relative
entropy of entanglement for given values of (a) negativity and (b)
nonlocality. It is seen that the maximal difference in the REE
between the optimal states and pure states is equal to (a)
$\Delta=\Delta E_R(N') = 0.0391$ at the negativity $N'=0.154$ and
(b) $\Delta=\Delta E_R(B') = 0.404$ at the nonlocality $B'=0$.
Pure states become more entangled than the Horodecki states
$\RHO{H}(p)=\RHO{A}(\frac12,p)$ for (a) $N>N(\rho_1)=0.377$ and
(b) $B>B(\rho_6)=0.586$, while pure states become optimal for (a)
$N>N_0\equiv N(\rho_2)=0.527$ and (b)  $B>B_0\equiv
B(\rho_4)=B(\rho_5)=0.817.$ The states $\rho_n$ (for $n=1,...,6)$
are specified in Table~\ref{tab:points}.}
\end{figure}
In Fig.~\ref{fig:1}, we have presented the entanglement and
nonlocality measures for $10^6$ randomly generated two-qubit
states using the Monte Carlo simulation. It is seen that pure
states are maximally entangled in terms of the REE for an
arbitrary concurrence as shown in Fig.~\ref{fig:1}(a).
However, pure states are not always maximally entangled for
the cases shown in Figs.~\ref{fig:1}(b) and~\ref{fig:1}(c).
Namely, the red regions correspond to mixed states having the
REE higher than that for pure states for given negativity
[Fig.~\ref{fig:1}(b)] and nonlocality [Fig.~\ref{fig:1}(c)].

This is a counterintuitive result, so to have a deeper
physical meaning of the states corresponding to the red
regions in Fig.~\ref{fig:1}, let us analyze the loss of
entanglement between two qubits initially in a pure state. In
particular, by taking one of the Bell states $\ket{\psi^\pm}
= (\ket{01}\pm\ket{10})/\sqrt{2}$ or $\ket{\phi^\pm} =
(I\otimes\sigma_1)\ket{\psi^\pm}$ as an initial state, one
can achieve any degree of entanglement between the two qubits
by coupling them to their environment(s). Note, however, that
by coupling qubits to separated and uncorrelated
environments, we can only decrease the entanglement, since
any local operation cannot increase it. Damping can occur in
various ways, but we focus on states generated from pure
states by two prototype damping models (channels), for which
the ranges of the REE for a given value of the nonlocality is
shown in Fig.~\ref{fig:2}. We show in detail that these
damping channels provide us with the mixed states with
extreme values of the REE for a fixed value of another
entanglement or nonlocality measures.

In this section we analyze the effect of the prototype damping
channels on  pure states of the form
\begin{equation}
|\psi_{\alpha}\rangle=\sqrt{\alpha}|01\rangle+\sqrt{1-\alpha}|10\rangle,\label{psi_in}
\end{equation}
where $0\le \alpha\le 1$. When analyzing the entanglement
measures, any pure two-qubit state, as given above
Eq.~(\ref{B_pure_state}), can be equivalently expressed by means
of the Schmidt decomposition, in the form given by
Eq.~(\ref{psi_in}). Although damping of these pure states can lead
to inequivalent states, for simplicity, we confine our analysis
only to the input states $|\psi_{\alpha}\rangle$.

A convenient description of damping can be given in terms of the
Kraus operators $E_{i}$ (specified below). Two-side damping of the
state $\rho_{{\rm in}} (\alpha)
=|\psi_{\alpha}\rangle\langle\psi_{\alpha}|$ leads to the
following output state
\begin{equation}
\rho(\alpha,q_{1},q_{2})=\sum_{i,j}[E_{i}(q_{1})\otimes
E_{j}(q_{2})]\rho_{{\rm in}}(\alpha)[E_{i}^{\dagger}(q_{1})\otimes
E_{j}^{\dagger}(q_{2})],\label{N03}
\end{equation}
where the Kraus operators satisfy the normalization relation
$\sum_{i}E_{i}^{\dagger}(q)E_{i}(q)=I$. In the special case of the
one-side damping (say $q_{2}=0$), Eq.~(\ref{N03}) reduces to
\begin{equation}
\rho(\alpha,q_{1}) =\sum_{i}[E_{i}(q_{1})\otimes I]\rho_{{\rm
in}}(\alpha)[E_{i}^{\dagger}(q_{1})\otimes I].\label{N05}
\end{equation}

\subsection{Amplitude-damping channel}

The Kraus operators for the single-qubit amplitude-damping channel
(ADC) are the following~\cite{Nielsen}:
\begin{equation}
E_{0}(q_{i})=|0\rangle\langle0|+\sqrt{p_{i}}|1\rangle\langle1|,\quad
E_{1}(q_{i})=\sqrt{q_{i}}|0\rangle\langle1|,\label{Kraus_adc}
\end{equation}
where $q_{i}\equiv 1-p_{i}$ is the amplitude-damping coefficient.
One can find that the pure state $|\psi_{\alpha}\rangle$ after passing
through the ADC is changed into the following mixed
state
\begin{equation}
\RHO{ADC}(\alpha,q_{1},q_{2})=p|\psi_{\alpha'}\rangle\langle\psi_{\alpha'}|
+q|00\rangle\langle00|\equiv\RHO{A}(\alpha',p) , \label{rho_adc}
\end{equation}
where the  effective damping constant reads  as $q\equiv
1-p=\alpha q_{2}+(1-\alpha)q_{1}$ and $|\psi_{\alpha'}\rangle$ is
given by Eq.~(\ref{psi_in}) but for $\alpha'=\alpha
p_{2}/[\alpha p_{2}+(1-\alpha)p_{1}]$, which can take values from
0 to $\alpha$, where $\alpha'=\alpha$  is achieved for the
symmetric two-side ADC, i.e., when $q_1=q_2$. The state given by
Eq.~(\ref{rho_adc}) is sometimes referred to as the generalized
Horodecki state~(see, e.g., Refs.~\cite{Miran08a,Miran08b}).
Hereafter, we refer to the discussed states $\RHO{A}(\alpha',p)$
as the amplitude-damped states. The range of the REE for a given
nonlocality for the ADC state, given by Eq.~(\ref{rho_adc}),  is
shown in Fig.~\ref{fig:2}.

In the special case, when  $\alpha p_{2}=(1-\alpha)p_{1}$,
which implies $\alpha'=1/2$,  the ADC reduces to the Horodecki
state (see e.g. Ref.~\cite{Sahin07}), i.e.,
\begin{equation}
\RHO{H}(p)=\RHO{A}(\textstyle{\frac12},p)=p|\psi^{+}\rangle\langle\psi^{+}|
+q|00\rangle\langle00|, \label{rho_h2}
\end{equation}
as given by Eq.~(\ref{rho_H}). These states can be obtained from
the initial Bell state $|\psi^{+}\rangle$ subjected to the
symmetric two-side ADC, but they can be also obtained by asymmetric two-side ADC (or even the one-side ADC) of
the initially non-maximally entangled pure state
$|\psi_{\alpha}\rangle$.

We find that the nonlocality of the ADC
state for any $\alpha$ is given by
\begin{equation}
B(\RHO{A})=\sqrt{\max\{0,\max[x,(1-2p)^{2}]+x-1\}},\label{B_ADC}
\end{equation}
where $x  =  4 (1-\alpha' ) \alpha'  p^2$. By comparison, the
negativity reads as
\begin{equation}
N(\RHO{A})=\sqrt{(1-p)^2+x}-(1-p),\label{N_ADC}
\end{equation}
and the concurrence is  $C(\RHO{A})=\sqrt{x}$. By contrast to
these simple formulas, there is no analytical formula for the
REE for $\RHO{A}$ for arbitrary $\alpha$ and $p$, except some
special cases. For example,  the REE can be calculated
analytically, according to Eq.~(\ref{REE_H}), for the
Horodecki states, given by Eq.~(\ref{rho_h2}).

Our numerical calculations shown in Fig.~\ref{fig:1}(c) indicate
that the Bell-diagonal states are likely to be the lower bound for
the REE vs nonlocality. Thus, let us analyze whether the ADC
states can be diagonal in the Bell basis. By denoting the Bell
states as follows $|\beta_{1}\rangle=|\psi^{-}\rangle$,
$|\beta_{2}\rangle=|\psi^{+}\rangle$,
$|\beta_{3}\rangle=|\phi^{-}\rangle$, and
$|\beta_{4}\rangle=|\phi^{+}\rangle$, we find that the ADC state can be
given in the Bell basis as:
\begin{eqnarray}
\RHO{ADC}(\alpha,q_{1},q_{2})  =r_{-}
|\beta_{1}\rangle\langle\beta_{1}|
+r_{+}|\beta_{2}\rangle\langle\beta_{2}|+r|\beta_{3}\rangle\langle\beta_{3}|
\nonumber
\\
 +r|\beta_{4}\rangle\langle\beta_{4}| +(t|\beta_{2}\rangle\langle\beta_{1}|
 +r|\beta_{4}\rangle\langle\beta_{3}|+{\rm H.c.}),\hspace{7mm} \label{N17c}
\end{eqnarray}
where $t=\alpha(1-q_{2})+r-\frac{1}{2}$ ;
$r_{\pm}=\frac{1}{2}[1-2r\pm C(\RHO{A})],$ and
$r=\frac{1}{2}[(1-\alpha)q_{1}+\alpha q_{2}]$. So, this state is
diagonal in the Bell basis only if $\alpha=0$ and 1. Note that for
$\alpha=1/2,$ one has  $r_{-}=\frac{1}{4}(p_{1}-p_{2})^2$, which
vanishes for $p_{1}=p_{2},$ but another off-diagonal term
$r=\frac{1}{4}(q_{1}+q_{2})$ vanishes only if there is no damping
at all. This shows (as also confirmed numerically in
Fig.~\ref{fig:2}) that the ADC states do not have the minimum of
the REE for a given nonlocality except only two points for $B=0$
and 1. By contrast, Figs.~\ref{fig:1}(c) and~\ref{fig:2} show that
the ADC states are likely to be the upper bound of the REE vs
nonlocality. This observation is confirmed analytically in
Sec.~\ref{sec:KKT}.

\subsection{Phase-damping channel}

The Kraus operators for the single-qubit phase-damping channel
(PDC) can be given by~\cite{Nielsen}:
\begin{equation}
E_{0}(q_{i})=|0\rangle\langle0|+\sqrt{p_{i}}|1\rangle\langle1|,\quad
E_{1}(q_{i})=\sqrt{q_{i}}|1\rangle\langle1|,\label{Kraus_pdc}
\end{equation}
where $q_i=1-p_{i}$ is the phase-damping coefficient. Thus, one
can find that the pure state $|\psi_{\alpha}\rangle$ after being
transmitted through the PDC is changed into the following mixed
state
\begin{eqnarray}
\RHO{PDC}(\alpha,q_{1},q_{2})&=&\alpha|01\rangle\langle01|
+(1-\alpha)|10\rangle\langle10|
\nonumber \\ &&+y(|01\rangle\langle10|+|10\rangle\langle01|),
\end{eqnarray}
where $y=\sqrt{\alpha(1-\alpha)p_{1}p_2}.$ This state can be also
given as
\begin{equation}
\RHO{PDC}(\alpha,q_{1},q_{2})=
p''|\psi_{\alpha''}\rangle\langle\psi_{\alpha''}|+q''|01\rangle\langle01|,
\label{rho_pdc}
\end{equation}
where the  effective damping constant can be defined as
$q''\equiv 1-p''=\alpha(1-p_{1}p_{2})$ and the pure state
$|\psi_{\alpha''}\rangle$ is  given by Eq.~(\ref{psi_in}) but
for $\alpha''=\alpha p_{1}p_{2}/p''$, which takes the values
$0\le \alpha''\le \alpha$.

We find that the nonlocality, negativity and concurrence for the
PDC state are equal to each other and are given by
\begin{equation}
B(\RHO{PDC})=N(\RHO{PDC})=C(\RHO{PDC})=2y.
\label{B_PDC}
\end{equation}
The range of the REE for a given nonlocality for the PDC state,
given by Eq.~(\ref{rho_pdc}),  is shown in Fig.~\ref{fig:2}. Note
that it is unlikely that there is an analytical formula for the
REE for the PDC states for arbitrary parameters $\alpha,\;q_{1}$,
and $q_{2}$. However, the REE can be found in some special cases.
For example, if $\alpha=[1+p_{1}p_{2}]^{-1}\equiv\alpha'$ then
Eq.~(\ref{rho_pdc}) reduces to
\begin{eqnarray}
\RHO{PDC}(\alpha',q_{1},q_{2})\equiv\RHO{V}=2(1-\alpha')|\psi^{+}\rangle\langle\psi^{+}|
\nonumber \\ +(2\alpha'-1)|01\rangle\langle01|,\label{N16}
\end{eqnarray}
which means that $|\psi_{\alpha'}'\rangle$ becomes the Bell state
$|\psi^{+}\rangle$.  For this state, one finds that
\begin{equation}
B'\equiv B(\RHO{V})=C(\RHO{V})=N(\RHO{V})=2(1-\alpha').\label{N17}
\end{equation}
The state $\RHO{V}$, the same as pure states, reaches the upper
bound for the negativity versus concurrence~\cite{Miran04a}. The
CSS for $\RHO{V}$ reads as
$\SIGMA{V}=(1-B'/2)|01\rangle\langle01|+(B'/2)|10\rangle\langle10|$.
Thus, the REE can simply be given by:
\begin{equation}
E_{R}(\RHO{V})=h(\textstyle{\frac12}B')-h\{\textstyle{\frac12}[\sqrt{(1-B')^{2}+(B')^{2}}+1]\}.\label{N18}
\end{equation}Now, we show that the minimum of the REE vs nonlocality can be
reached by the PDC states. In the Bell basis, the PDC state is
given by
\begin{eqnarray}
\RHO{PDC}(\alpha,q_{1},q_{2})=(\textstyle{\frac{1}{2}}-y)|\beta_{1}\rangle\langle\beta_{1}|
+(\textstyle{\frac{1}{2}}+y)|\beta_{2}\rangle\langle\beta_{2}|\nonumber \\
 +(\alpha-\textstyle{\frac{1}{2}})(|\beta_{1}\rangle\langle\beta_{2}|+|\beta_{2}\rangle\langle\beta_{1}|),
\hspace{5mm}
\end{eqnarray}
which clearly becomes diagonal for initial Bell states (i.e., for
$\alpha=1/2$), given by Eq.~(\ref{rho_dc2}) for $q\equiv
1-p=1/2-y$ (see e.g. Ref.~\cite{Sahin07}). The REE for the state
$\RHO{D}$ is given by Eq.~(\ref{REE_B}). Hereafter, we consider
only the case when $\alpha = 1/2$ and use the shorthand notation
$\RHO{D}\equiv\RHO{D2}$.
\section{Karush-Kuhn-Tucker conditions for the REE vs nonlocality}
\label{sec:KKT}

In this section, we derive the KKT conditions, as a generalization
of the method of Lagrange multipliers, in order to find the states
with extremal values of the REE for a fixed value of the
nonlocality.

Thus, let us consider the following Lagrange function:
\begin{eqnarray}
{\cal L}&=&\hbox{Tr(}\rho \ln \rho) - \hbox{Tr} (\rho \ln
\sigma)
+ l\hbox{[Tr(} \rho {\cal B}_{\rm CHSH})-\beta]\nonumber\\
&& -\hbox{\;Tr(}X\rho)+\lambda(\hbox{Tr}\rho-1),
\end{eqnarray}
where $l$, $\lambda$, and $X$ are Lagrange multipliers and
$\beta=2\sqrt{M(\rho)}$. For a small deviation of
$\rho\rightarrow\rho+\Delta$, we have
\begin{eqnarray}
{\cal L}&\rightarrow&{\cal
L}+\hbox{Tr(}\Delta\ln\rho)-\hbox{Tr(}\Delta\ln\sigma)
\nonumber\\&& +\;\hbox{Tr}\Big(\rho\int_0^\infty
\frac{1}{\rho+z}\Delta\frac{1}{\rho+z}{\rm d}z\Big)
\nonumber\\&& +\;l\hbox{Tr(}\Delta{\cal B}_{\rm CHSH})
-\hbox{Tr(}\Delta X)+\lambda\hbox{Tr}\Delta \cr
&\rightarrow&{\cal
L}+\hbox{Tr[}\Delta(\ln\rho-\ln\sigma+P\nonumber\\&&
+\;l{\cal B}_{\rm CHSH}-X+\lambda)],
\end{eqnarray}
where $P$ is a projector on the support space of $\rho$. Thus, the
KKT conditions on the parameters $l$, $\lambda$, and $X$ are the
following:
\begin{subequations}
\begin{eqnarray}
&&\ln\rho-\ln\sigma+P+l{\cal B}_{\rm
CHSH}-X+\lambda=0,\label{KKT1} \\
&&
 \hbox{~~~}
X\ge0, \hbox{~~~} \hbox{Tr}(X\rho)=0.
\end{eqnarray}
\end{subequations}
From the condition~(\ref{KKT1}), we obtain
\begin{eqnarray}
0&=&\hbox{Tr(}\rho\ln\rho)-\hbox{Tr(}\rho\ln\sigma)+\hbox{Tr(}\rho
P)\nonumber\\&&+\;l\hbox{Tr(}\rho{\cal B}_{\rm
CHSH})-\hbox{Tr(}\rho X)+\lambda\hbox{Tr}\rho\nonumber\\&=&
\gamma E_R(\rho)+1+l\beta+\lambda,\quad
\end{eqnarray}
where $\gamma=1/\log_2 e$ and $e$ is the base of the natural
logarithm. Thus, $\lambda=-\gamma E_R(\rho)-1-l\beta$. Now, we can
rewrite Eq.~(\ref{KKT1}) and obtain the simplified KKT conditions:
\begin{subequations}
\begin{eqnarray}
0&=&\ln\rho-\ln\sigma+P+l{\cal B}_{\rm
CHSH}
-X\nonumber\\&&-\gamma E_R(\rho)-1-l\beta,\\&& \hbox{~~~} X\ge0,
\hbox{~~~} \hbox{Tr}(X\rho)=0. \label{ExtremalCondition}
\end{eqnarray}
\end{subequations}
One can easily confirm that the condition
\begin{equation}
P\big[\ln\rho-\ln\sigma+l{\cal B}_{\rm CHSH}-\gamma
E_R(\rho)-l\beta\big]P=0
\end{equation}
is a new KKT condition since $PX=0$.

We search for the boundary states among the rank-2 states. We have
strong numerical evidence (see points for randomly generated
density matrices in Fig.~\ref{fig:1}) implying that the extreme
two-qubit states can be found within the rank-2 class of mixed
states. Thus, let us now consider the case where $\rho$ is a
rank-2 mixed state, i.e., $\rho=\lambda_1|e_1\rangle\langle e_1|+
\lambda_2|e_2\rangle\langle e_2|$, where $\lambda_i$ are nonzero
eigenvalues of $\rho$, and $|e_i\rangle$ are the corresponding
eigenstates. From Eq.~(\ref{ExtremalCondition}), we have
\begin{subequations}
\begin{eqnarray}
\langle e_1|\ln \sigma|e_2\rangle &=&
l \langle e_1|{\cal B}_{\rm CHSH}|e_2\rangle, \label{Condition0} \\
\gamma E_R(\rho)+l \beta &=& \ln\lambda_1 -\langle e_1|\ln
\sigma|e_1\rangle \nonumber\\&&+\; l \langle e_1|{\cal
B}_{\rm CHSH}|e_1\rangle, \label{Condition1}
\end{eqnarray}
\end{subequations}
from which $l$ is determined. Then, there must exist $X$ such that
\begin{subequations}
\begin{eqnarray}
X&=&\ln\rho-\ln\sigma+P+l{\cal B}_{\rm CHSH}\nonumber\\&&
-[\gamma E_R(\rho)+1+l\beta]\ge0, \hbox{~}\label{condition2a}\\
\hbox{Tr(}X\rho)&=&0. \label{condition2}
\end{eqnarray}
\end{subequations}
The KKT conditions, given by
Eqs.~(\ref{Condition0})--(\ref{condition2}), are the necessary
conditions in searching for the boundary (extreme) states among
rank-2 mixed states.

\subsection{Lower bound of REE vs nonlocality}

Here, we show the KKT conditions are satisfied by the
Bell-diagonal states, given by Eq.~(\ref{rho_dc2}), as obtained by
the phase damping of the Bell states $\ket{\psi^+}$ or,
equivalently, $\ket{\psi^-}$ assuming $p \leftrightarrow 1-p$.

For these states the CSS is given by
\begin{eqnarray}
\SIGMA{D}&=&\frac{1}{2}(|\psi^+\rangle\langle\psi^+|
+|\psi^-\rangle\langle\psi^-|),
\end{eqnarray}
 and the Bell-CHSH operator is
\begin{eqnarray}
{\cal B}_{\rm
CHSH}^{\mathrm{(D)}}&=&\eta\big[-\sigma_3\otimes\sigma_3+(2p-1)\sigma_1\otimes\sigma_1\big]
\nonumber \\ &=& 2\eta \big[
p(|\psi^+\rangle\langle\psi^+|-|\phi^-\rangle\langle\phi^-|)
\nonumber\\&&+\;(1-p)(|\psi^-\rangle\langle\psi^-|-|\phi^+\rangle\langle\phi^+|)\big],\;
\end{eqnarray}
where $\eta=2/\sqrt{1+(2p-1)^2}$. So, by applying
Eqs.~(\ref{MaxBellOp}) and~(\ref{BIV}), one can find the
simple expression, given by Eq.~(\ref{B_D2}), for the
nonlocality.

Since it holds $\langle e_1|\ln\SIGMA{D}|e_2\rangle= \langle
e_1|{\cal B}_{\rm CHSH}^{\mathrm{(D)}}|e_2\rangle=0$,
Eq.~(\ref{Condition0}) is satisfied for any $l$. From
Eq.~(\ref{Condition1}), $l$ is determined through
\begin{equation}
\ln p +1+2\eta pl=\gamma \ER{D}+l\beta,
\end{equation}
where $\beta=2\sqrt{ 1+(2p-1)^2}$ and the REE is given by
\begin{equation}
\ER{D} \equiv E_R(\RHO{D})=1-h(p),
\end{equation}
where $h(p)$ is the binary entropy defined below Eq.~(\ref{N35b}).
Then, the KKT conditions are satisfied if
\begin{eqnarray}
X&=&-[\gamma \ER{D}+1+l\beta+2\eta l(1-p)]|\phi^+\rangle\langle\phi^+|\nonumber\\
&&-[\gamma \ER{D}+1+l\beta+2\eta
lp]|\phi^-\rangle\langle\phi^-|\ge 0
\end{eqnarray}
for all values of $p$. Hence, the Bell-diagonal states can yield
the extreme (in this case minimum) value of the REE for a given
value of the nonlocality. This conclusion is confirmed by our
Monte Carlo simulations (see Fig.~\ref{fig:1}).

\subsection{Upper bound of REE vs nonlocality}

Here, we show that the KKT conditions are also satisfied by
the amplitude-damped states $\RHO{A}(\alpha,p)$, given by
Eq.~(\ref{rho_adc}), for properly chosen parameters $\alpha$
and $p$. These MEMS having the highest REE for a given
nonlocality are shown by curves $A''$ in Figs.~\ref{fig:1}(c)
and~\ref{fig:3}(b).

The support space of $\RHO{A}(\alpha,p)$ is given by the
eigenvectors $|e_1\rangle = |00\rangle$ and  $|e_2\rangle =
|\psi_{\alpha}\rangle$. The corresponding CSS is given
as~\cite{Miran08b}:
\begin{eqnarray}
\SIGMA{A}(\alpha,p) &=&
R_1|00\rangle \langle 00| +R_4|11\rangle \langle 11|\notag \\
&&+\lambda_{+}|\lambda_{+}\rangle \langle \lambda_{+}|
+\lambda_{-}|\lambda_{-}\rangle \langle \lambda_{-}|,
\label{CSSGH}
\end{eqnarray}
where
\begin{eqnarray}
\lambda_{\pm}&=&\frac{1}{2}\left[R_2+R_3\pm\sqrt{(R_2-R_3)^2+4R_1R_4}
\right]\notag,\\
|\lambda_{\pm}\rangle&=&
\Lambda_{\pm}\left[(\lambda_{\pm}-R_3)|01\rangle+\sqrt{R_1R_4}|10\rangle\right],
\end{eqnarray}
normalized by
$\Lambda_{\pm}=\left[(\lambda_\pm-R_3)^2+R_1R_4\right]^{-1/2}$. To
calculate the CSS we have to compute
\begin{subequations}
\begin{eqnarray}
R_2 &=&\frac{1}{4}\left[1+3(1-p)+2p\alpha-4R_1 - \sqrt{\delta}\right],\nonumber\\
R_4 &=& R_1 - 1 + p, \\
R_3 &=& 1 - \sum_{i=1,2,4}R_i,\nonumber\\
\delta &=& (4-3p)^2 - 4\alpha(1-\alpha)p^2-8R_1(2-p)\nonumber\\
&&+16\sqrt{R_1(R_1-1+p)p^2\alpha(1-\alpha)},
\end{eqnarray}
\end{subequations}
where $R_1$ is obtained by solving
\begin{eqnarray} \label{problem}
\alpha p &=& R_2 + 2R_4(R_2^2-R_2R_3+2R_1R_4)/z^2\nonumber \\
    &&+ 2R_4(R_2-R_3)/(Lz),
\end{eqnarray}
where $z=\sqrt{(R_2-R_3)^2+4R_1R_4}$ and $L=\ln(R_2+R_3-z) -
\ln(R_2+R_3+z)$. The Bell-CHSH operator in this case reads as
\begin{eqnarray}
\label{BIVGH} {\cal B}_{\rm CHSH}^{\mathrm{(A)}}
= \begin{cases} \eta_1\left[(1-2p)\sigma_3^{\otimes2} + 2p\sqrt{(1-\alpha)\alpha}\sigma_1^{\otimes2} \right]\\
                \qquad \mbox{     if } 4p^2(1 - \alpha)\alpha - (1 - 2p)^2 <0,\\
                \eta_2p\sqrt{(1-\alpha)\alpha}(\sigma_1^{\otimes2} +\sigma_2^{\otimes2})\\
                \qquad\mbox{     otherwise,}
\end{cases}
\end{eqnarray}
where $\eta_1=2/\sqrt{(1-2p)^2+4p^2\alpha(1-\alpha)}$,
$\eta_2=2/\sqrt{8p^2(1-\alpha)\alpha}$. Thus, by applying
Eqs.~(\ref{MaxBellOp}) and~(\ref{BIV}), we find that the
nonlocality is given by Eq.~(\ref{B_ADC}). Since we fix  the value
of $\beta$ or equivalently  $B\equiv B(\RHO{A})$, we can express $p$ in
terms of $B$ and  parameter $\alpha$ in the following way
\begin{eqnarray}
\alpha=\begin{cases} \frac{1}{4p}\left(2p - \sqrt{2(2p^2 - B^2 - 1)}\right)\\\qquad\mbox{ if } 2\sqrt{2+2B^2}\leq 4p \leq 2+\sqrt{2+2B^2},\\
                \frac{1}{2p}\left(p - \sqrt{5p^2 - 4p - B^2}\right)\\\qquad\mbox{ if } 2+\sqrt{2+2B^2} <4p\leq 4.\end{cases}
\end{eqnarray}
We can easily check by using Eqs.~(\ref{BIVGH}) and~(\ref{CSSGH})
that the condition, given by Eq.~(\ref{Condition0}), is always
satisfied. Thus, the condition, given by Eq.~(\ref{Condition1}),
is also satisfied. To check the remaining conditions, we need an
explicit expression for the REE and multiplier $l$, which is
equivalent to solving Eq.~(\ref{problem}). This equation contains
logarithms and can be easily solved only in special cases such as,
e.g., the Horodecki states ($\alpha=1/2$). Nevertheless, our
numerical analysis reveals that for $\RHO{A}(\alpha,p)$, for which
the REE reaches the largest value for a given value of
nonlocality, Eq.~(\ref{condition2}) is satisfied for the following
coefficients of the amplitude-damped states:
\begin{subequations}
\begin{eqnarray}
p &=& {\begin{cases} \frac{1}{4}\left(2 + \sqrt{2 +
2B^2}\right)\qquad\mbox{ if } B <B_0,\\
1\hspace{3.3cm}\mbox{ if } B >B_0,
              \end{cases} }\\
1&\geq&p\geq \frac{1}{4}\left(2 + \sqrt{2 +
2B_0^2}\right)\mbox{ if } B = B_0,\\
\alpha &=& \frac{1}{2p}\left(p - \sqrt{5p^2 - 4p - B^2}\right),
\end{eqnarray}
\end{subequations}
where $B_0=0.816\,86$ [see Fig.~\ref{fig:3}(b)].

Thus,  by comparing the REEs for a given nonlocality for the
optimal mixed states (denoted by superscript A'') and pure states,
we can conclude that
\begin{eqnarray}
\ER{A''}(B)>\ER{P}(B) & \quad & {\rm for}\;0<B<B_0,\label{N85a}\\
\ER{A''}(B)<\ER{P}(B) & \quad & {\rm for}\; B_0<B<1,\nonumber
\end{eqnarray}
where $B_0=B(\rho_{4})=B(\rho_{5})=0.816\,86$ and
$\ER{A''}(B_{0})=\ER{P}(B_{0})=0.7445$ [see Table~I and
Fig.~3(b)]. On the other hand, by comparing the REE for a given
negativity for the optimal mixed states (denoted by superscript
A') and pure states, it holds~\cite{Miran08b}:
\begin{eqnarray}
\ER{A'}(N)>\ER{P}(N) & \quad & {\rm for}\;0<N<N_0,\label{N86a}\\
\ER{A'}(N)<\ER{P}(N) & \quad & {\rm for}\; N_0<N<1,\nonumber
\end{eqnarray}
where $N_0=N(\rho_{2})=0.5271$ and
$\ER{A'}(N_{0})=\ER{P}(N_{0})=0.3847$ [see Table~I and
Fig.~\ref{fig:3}(a)]. As a reminder, superscripts A' (A'')
correspond to the amplitude-damped states $\RHO{A'}$
($\RHO{A''}$) optimized for the REE vs negativity
(nonlocality). As shown in Fig.~\ref{fig:1}, the states
$\RHO{A'}$ and $\RHO{A''}$ are, in general, inequivalent. The
maximal differences, as defined by Eq.~(\ref{Delta}), are
equal to $\Delta E_R(N)=0.0391$ for $N=0.1540$, and  $\Delta
E_R(B)=0.4040$ for $B=0$. It is seen that ranges of mixed
states, which are more entangled than pure states, are much
smaller for the REE vs negativity in comparison to  the REE
vs nonlocality.

\section{Conclusions}

We studied the relation between the relative entropy of
entanglement $E_R$ and the nonlocality measure $B$ corresponding
to a degree of the violation of the Bell-CHSH inequality in
two-qubit systems. We found states of the extremal value of $E_R$
for a given value of $B$. We showed that the obtained states
satisfy the Karush-Kuhn-Tucker conditions, derived in
Sec.~\ref{sec:KKT}, as well as they provide a boundary for the REE
values obtained by our Monte Carlo simulations presented in
Fig.~\ref{fig:1}.

We demonstrated that mixed states can be more entangled in terms
of $E_R$ than pure states if the nonlocality $B\in(0,0.82)$ and
the maximal difference between these REEs is $\Delta E_R=0.4$ as
shown in Fig.~\ref{fig:3}(b). As discussed in
Ref.~\cite{Miran08b}, $E_R$ as a function of the negativity $N$
can also exhibit this property but (i) the maximal difference
$\Delta E_R$ is one order smaller (i.e., $\Delta E_R=0.039$) and
(ii) mixed states are optimal for a shorter range of the
negativity, i.e., $N\in(0,0.53)$, as presented in
Fig.~\ref{fig:3}(a). For appropriate comparison, we normalized $B$
to be equal to $N$ for any two-qubit pure state.

We showed that these maximally entangled mixed states can be
obtained from pure two-qubit entangled states by locally
subjecting one or both qubits (of the entangled pair) to the
amplitude-damping channel, while the minimally entangled states
can be generated from pure states by subjecting  one or both
qubits to the phase-damping  channel. We found that the
amplitude-damped states (yellow and green areas in
Fig.~\ref{fig:2}) are more entangled than the phase-damped states
(blue and green areas) for a given nonlocality $B<0.8169$
(corresponding to the nonlocality of the states $\rho_4$ and
$\rho_5$ shown in Fig.~\ref{fig:1} and Table~\ref{tab:points}).

However, for values $B>0.5856$ (point $\rho_6$) there exists
a range of states, obtained either by the amplitude or phase
damping, which have the same value of $E_R$ for a given $B$
(green area in Fig.~\ref{fig:2}). Moreover, we found that the
upper bound on the REE of the phase-damped states (blue and
green areas in Fig.~\ref{fig:2}) for a given nonlocality is
provided by pure states (curve $P$ in Fig.~\ref{fig:2}).
However, pure states have the highest REE only for
$B>0.8169$.

Thus, we found that for a large range of the nonlocality, mixed
states can be more entangled than pure states, and, surprisingly,
these mixed states can by obtained by the ordinary amplitude
damping of pure states.

We note that two-qubit pure states are extremal (i.e., they
are on the \emph{lower} bound) for the concurrence $C$ vs
$B$, as shown in Ref.~\cite{Verstraete02}, and the negativity
$N$ vs $B$~\cite{unpublished} for arbitrary $B\in[0,1]$. This
means that all mixed states are more entangled than pure
states or, at least, the same entangled in terms of $C$ and
$N$ for a given $B$. The distinctive feature of the relation
of the REE vs $B$, is that pure states are extremal (i.e.,
they are on the \emph{upper} bound) in some range of $B$
only. Thus, in the other range of $B$, mixed states can be
both more and less entangled than pure states.

Finding relations between the REE and nonlocality (or other
measures of quantum correlations) is impeded by the lack of
an analytical formula for the REE for general two-qubit mixed
states. By contrast, such formulas are known for the
two-qubit concurrence and negativity. Thus, a related problem
of comparing these two quantities with the nonlocality is
much simpler.

We believe that these results can stimulate a further quest for
practical protocols of quantum information processing for which
mixed states are more effective than pure states.

\begin{acknowledgments}
We thank Dr. Satoshi Ishizaka for his insightful explanations.
This work was supported by the Polish National Science Centre
under Grants No. DEC-2011/03/B/ST2/01903 and No.
DEC-2011/02/A/ST2/00305. K.~B. gratefully acknowledges the support
by the Operational Program Research and Development for
Innovations--European Regional Development Fund (Project No.
CZ.1.05/2.1.00/03.0058) and the Operational Program Education for
Competitiveness--European Social Fund (Project No.
CZ.1.07/2.3.00/30.0041) of the Ministry of Education, Youth and
Sports of the Czech Republic.
\end{acknowledgments}

\end{document}